\documentclass[reprint,superscriptaddress,amsmath,amssymb,aps]{revtex4-1}

\usepackage{bm}
\usepackage{varwidth}
\usepackage{algpseudocode, algorithm}

\usepackage{natbib}
\usepackage{graphicx}
\usepackage{braket}
\usepackage{physics}
\usepackage{tabularx}
\usepackage{amsmath}
\usepackage[colorlinks=true, allcolors=blue]{hyperref}

\begin{document}
\preprint{APS/123-QED}
\title{Quantum Multiple Kernel Learning}

\author{Seyed Shakib Vedaie}
\affiliation{1QB Information Technologies (1QBit), Vancouver, BC, Canada}
\affiliation{Institute for Quantum Science and Technology, University of Calgary, AB, Canada}

\author{Moslem Noori}
\affiliation{1QB Information Technologies (1QBit), Vancouver, BC, Canada}

\author{Jaspreet S. Oberoi}
\affiliation{1QB Information Technologies (1QBit), Vancouver, BC, Canada}
\affiliation{School of Engineering Science, Simon Fraser University, Burnaby, BC, Canada}

\author{Barry C.\ Sanders}
\affiliation{Institute for Quantum Science and Technology, University of Calgary, AB, Canada}

\author{Ehsan Zahedinejad}
\affiliation{1QB Information Technologies (1QBit), Vancouver, BC, Canada}

\begin{abstract}

Kernel methods play an important role in machine learning applications due to their conceptual simplicity and superior performance on numerous machine learning tasks. Expressivity of a machine learning model, referring to the ability of the model to approximate complex functions, has a significant influence on its performance in these tasks. One approach to enhancing the expressivity of kernel machines is to combine multiple individual kernels to arrive at a more expressive combined kernel. This approach is referred to as multiple kernel learning (MKL). In this work, we propose an MKL method we refer to as quantum MKL, which combines multiple quantum kernels. Our method leverages the power of deterministic quantum computing with one qubit (DQC1) to estimate the combined kernel for a set of classically intractable individual quantum kernels. The combined kernel estimation is achieved without explicitly computing each individual kernel, while still allowing for the tuning of individual kernels in order to achieve better expressivity. Our simulations on two binary classification problems---one performed on a synthetic dataset and the other on a German credit dataset---demonstrate the superiority of the quantum MKL method over single quantum kernel machines.

\end{abstract}
\maketitle

\section{Introduction}
Noisy, intermediate-scale quantum (NISQ) technologies have enabled the demonstration of~\emph{quantum supremacy}---the first milestone in quantum information processing~\cite{Pre18,AAB+19}.
The next milestone in quantum computing will be the development of applications that leverage the limited quantum resources offered by NISQ devices to demonstrate a quantum advantage for specific tasks. An example of such a task is the teaching of computers to learn from data---a fundamental problem in artificial intelligence. Despite recent progress in using quantum information processing for machine learning applications~\cite{BWP+17}, we are still in an exploratory phase in terms of our understanding of the potential capacity of quantum machine learning for solving real-world problems.

Currently, there exist several quantum machine learning algorithms that employ NISQ technologies. There are multiple examples within the area of quantum neural networks~\cite{KDT+20,FN18,SSP14} and some others in the area of quantum kernel methods~\cite{HCT+19,SK19}. The basic idea behind kernel methods is to use a kernel function as a measure of similarity between two data samples. This similarity measure is then used in various machine learning methods for solving classification or regression problems. In the age of complex deep learning methods, kernel methods~\cite{KSG+01}, despite their simplicity, are still one of the widely used tools in machine learning, especially when the dataset is small~\cite{HHA+06,ZL18}.

Recently, a quantum kernel machine has been proposed~\cite{GZO19} that uses a non-universal quantum computing model called deterministic quantum computing with one qubit (DQC1)~\cite{KL98}. Despite its non-universality, quantum algorithms that use DQC1 can efficiently solve a wide range of important classically intractable problems~\cite{PBL+04,SJ07,BS08}. To this date, all efforts to efficiently simulate DQC1 using classical computing resources have failed~\cite{DV07}, whereas its experimental realization across various quantum platforms has been successfully demonstrated~\cite{LBA+08,PMR+09,WHY+19,GTB+20}. The key idea in this method is to use the trace of a unitary operator as a kernel function to represent the similarity between two data samples in a high-dimensional quantum feature space~\cite{GZO19}. For a unitary operator whose trace is classically intractable to compute, DQC1 provides an exponential speed-up over any known classical algorithm to estimate the normalized trace. As a result, a quantum advantage in kernel estimation is achieved for classically intractable kernels.
 
One way to improve the efficacy of a machine learning model is to enhance its \emph{expressivity}, referring to the ability of the model to approximate complex functions. One of the ways to enhance the expressivity of kernel methods is to combine multiple kernels in order to arrive at a more expressive combined kernel~\cite{ME11}. This approach is called multiple kernel learning (MKL). In principle, each kernel in the combination can represent a unique notion of similarity for a specific subset of data features. Recent applications of MKL include anomaly detection~\cite{CRS+19}, heterogeneous data integration~\cite{MV17}, a recommendation system for heart disease diagnosis~\cite{MVP18}, and feature selection~\cite{XSX20}.

Our main contribution in this work is to propose a multiple kernel learning method called QMKL that leverages the power of DQC1 to efficiently estimate the combined kernel of a set of classically intractable quantum kernels. In its most general form, we consider the combination to be a weighted sum or product of more than one parameterized quantum kernels. In our method, the estimation of the combined kernel is achieved without explicitly computing each individual kernel while allowing for the tuning of the combined kernel to achieve better expressivity~\cite{YMT+18}. In contrast, tuning each individual parameterized quantum kernel can be performed via a variational quantum circuit~\cite{BLS+19} where the performance of the kernel-based model can be improved by varying the parameters of the underlying quantum circuit.

To show the efficacy of our quantum machine learning method, we consider two classification tasks: a synthetic dataset and a German credit dataset~\cite{DC17}. The results of our simulations show the superiority of QMKL over a single quantum kernel. Compared to single quantum kernel learning (SQKL), QMKL improves the average test accuracy by $21.69\%$ and $2.11\%$  for the synthetic dataset and the German credit dataset, respectively.

This work is structured as follows. In Section~\ref{sec:backg}, we provide a background on kernel methods and DQC1 and the connection between them. We also give an overview of the MKL approach. In Section~\ref{sec:qmkl}, we discuss our main contribution comprising different ways to combine quantum kernels using DQC1. Section~\ref{sec:simulation} lays out the details of the simulations we perform. We report our results in Section~\ref{sec:results} and discuss the efficacy of our proposed method on the benchmarking problems in Section~\ref{sec:discussion}, before giving our conclusion.

\section{Background}
\label{sec:backg}
The proposed QMKL method consists of several components. In what follows, we provide a brief background on each of these components comprising kernels methods, the DQC1 computing model, and its applications in kernel methods. We conclude the section by giving a summary of classical MKL.

\subsection{Kernel Methods}
Without loss of generality, here we explain  kernel methods for the case of classification problems in supervised machine learning.
In a classification problem, we are given a dataset $\mathcal{D}=\{\bm{x}_i, y_i\}_{i=1,\ldots, N}$ where $\bm{x}_i \in \mathbb{R}^p$, $y_i \in \{-1, +1\}$, and $N$ is the total number of data samples. Let us consider $\mathcal{X} \subset \mathbb{R}^p$ and $\mathcal{Y} \subset \{-1, +1\}$ as the sets that contain the feature data (each $\bm{x}_i$) and the label data (each $y_i$), respectively. The goal in such a learning task is to find a function $f: {\mathcal{X}} \rightarrow{\mathcal{Y}}$ on the pairs of feature-label data, such that,
given a set of unseen feature data, $f$ can predict the corresponding label data with a high probability.

\begin{figure}[ht]
    \centering
    \includegraphics[width=0.47\textwidth]{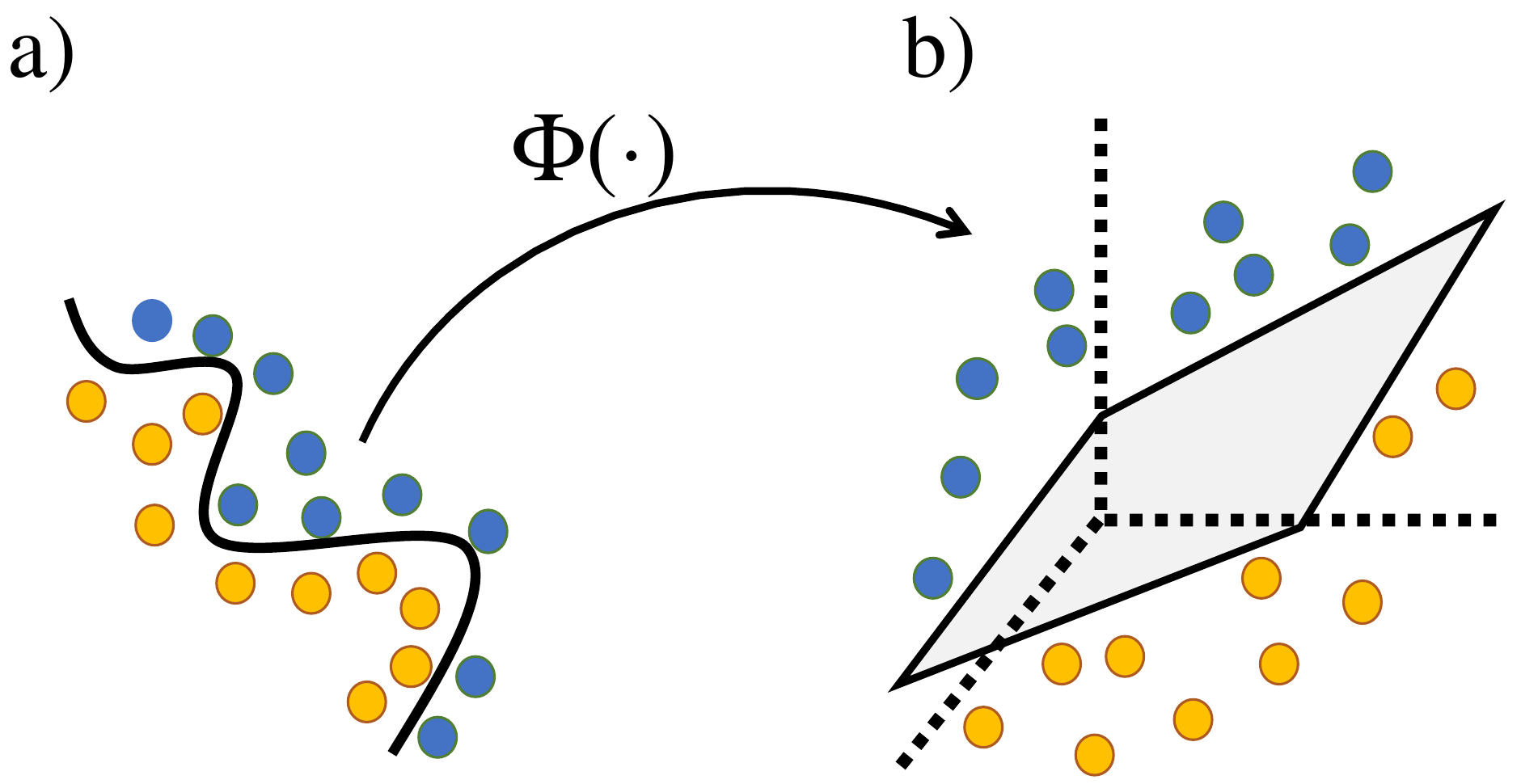}
    \caption{Example of the role of a kernel function for a) a binary classification task of blue vs. yellow classes. Using an explicit feature map, $\Phi(\cdot)$, one can transform the features of each data sample into a b) higher-dimensional feature space where the two classes become linearly separable. The kernel function is evaluated as the inner product of the two data points in the feature space.}
    \label{fig:kernel}
\end{figure}

To improve the separability of the classes in a dataset, we can transform them from their original space to a higher-dimensional Hilbert space, denoted by $\mathcal{H}$ (see Fig.~\ref{fig:kernel}). In machine learning terminology, $\mathcal{H}$ is often referred to as a~\emph{feature space}. Such a transformation is achieved by applying a feature map 
\begin{align}
    \label{eq:feature_map}
    \Phi:&\,\mathcal{X} \rightarrow{\mathcal{H}},\\ \nonumber
    & \bm{x}_ i\rightarrow{\Phi({\bm{x}_i}}).
\end{align}
By training a machine learning algorithm in $\mathcal{H}$ instead of $\mathbb{R}^p$, we aim to find the function~$f$ over the pairs $\{\Phi(x_i),y_i\}_{i=1,\ldots,N}$. The main bottleneck here is to explicitly calculate $\Phi$ for the data samples, as it becomes computationally challenging as the dimensionality of the feature space increases.

Kernel methods take advantage of the fact that the mapping of the feature data can be achieved without concern for the dimensionality of the feature space. The key idea behind kernel methods is that, in order to find the function $f$, one does not necessarily need to know the explicit form of feature map $\Phi$---only the overlap of the points in the feature space. This overlap represents a measure of similarity between the features of any two data samples and is called the kernel function, $k: \mathcal{X} \times \mathcal{X} \rightarrow{\mathbb{R}}$.

For two feature data $\bm{x}_i$ and $\bm{x}_j$, we have
\begin{equation}
    \label{eq:kernel_element}
    k(\bm{x}_i,\bm{x}_j):=\langle{\Phi(\bm{x}_i)},\Phi(\bm{x}_j)\rangle,
\end{equation}
where $\langle\cdot\,,\cdot\rangle$  denotes the inner product. Through the representer theorem~\cite{KW71,SHS01}, we can express the function $f$, which we call a base-learner, as the weighted sum of kernel functions:
\begin{equation} \label{eq:rep_theorem}
    f(\bm{x}) = \sum_{i = 1}^N \beta_i k(\bm{x}_i,\bm{x})\,,
\end{equation}
where $\beta_i\in \mathbb{R}$ is an element of the $N$-dimensional vector $\bm{\beta}$, the elements of which we call base-learner parameters.

\begin{figure*}[ht]
    \centering
    \includegraphics[width=0.6\textwidth]{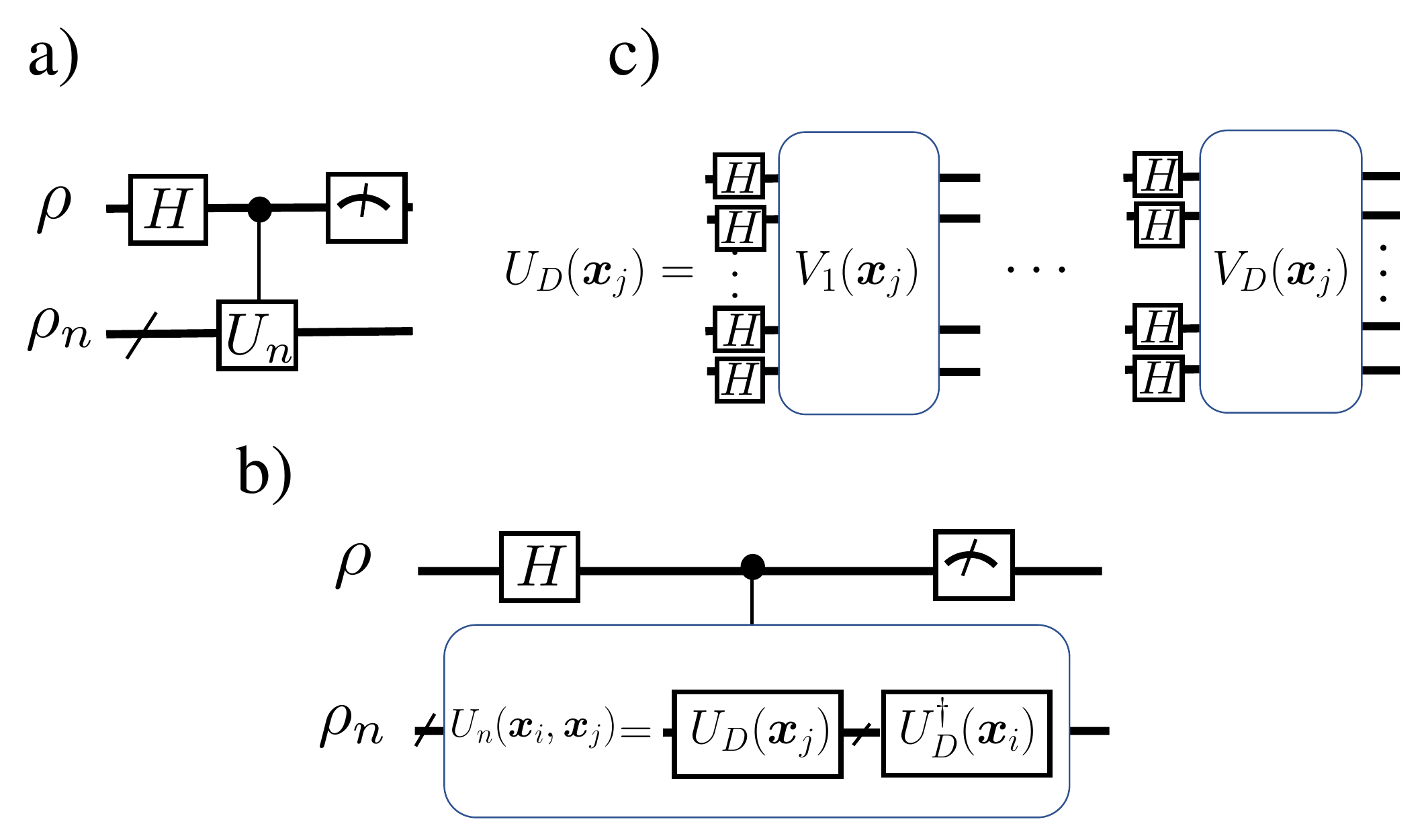}
    \caption{a) Trace estimation circuit for DQC1. The initial state of the control qubit and register qubits are denoted by $\rho$ and $\rho_n$, respectively. $H$ is the standard Hadamard gate and $U_n$ acts on the register qubits. b) DQC1 circuit for estimating the kernel function for the features of the two data samples $\bm{x}_i$ and $\bm{x}_j$. $U_D(\bm{x}_j)$ is called the encoding block. c) Example of an encoding pattern where each $V_i$ is a unitary operators defined by Eq.~\eqref{eq:encoding}.}
    \label{fig:DQC1}
\end{figure*}

\subsection{DQC1}
Figure~\ref{fig:DQC1}(a) shows the structure of the quantum circuit used in DQC1 for normalized trace estimation of a unitary operator. The circuit employs a single control qubit which is initialized in $\rho=\ketbra{0}{0}$ and $n$ register qubits which are initialized in a fully mixed state $\rho_n=\frac{\mathbb{I}_n}{2^n}$ where $\mathbb{I}_n$ is the $n$-qubit identity matrix. The unitary operators $H$ (here referring to a Hadamard gate) and $U_n$ act on the control qubit and register qubits, respectively. The initial state $\rho_\text{i}=\rho\otimes{\rho_n}$ of the quantum system evolves under the unitary operator $U=\ketbra{0}{0}\otimes{\mathbb{I}_n}+\ketbra{1}{1}\otimes{U_n}$.

Taking the partial trace over the register qubits, the final state of the control qubit, $\rho_\text{f}$, becomes
\begin{equation}
\label{eq:control_qubit_state}
    \rho_\text{f}=\frac{1}{2}
    \begin{bmatrix}
    1 & \text{tr}\left[\rho_nU^\dagger_n\right]\\
    \text{tr}\left[\rho_nU_n\right] & 1
\end{bmatrix},
\end{equation}
where $\text{tr}\left[\cdot\right]$ denotes the trace operator.
Considering that $\rho_n$ is a fully mixed state, the real (\text{Re}) and imaginary~(\text{Im}) parts of the normalized trace of $U_n$ can be estimated by measuring the expected value, denoted by $\left\langle {\cdot} \right\rangle$, of the Pauli-$X$ ($X$) and Pauli-$Y$ ($Y$) operators as follows:
\begin{align}
\label{eq:XY_estimation}
    \left\langle X \right\rangle_{\rho_{\text{f}}} & = \text{tr}\left[\rho_{\text{f}}X\right] = \text{Re}\left(\text{tr}\left[\rho_nU_n\right]\right) = \frac{1}{2^{n+1}} \text{Re}\left(\text{tr}\left[U_n\right]\right),\\ \nonumber
\left\langle Y \right\rangle_{\rho_{\text{f}}} & = \text{tr}\left[\rho_{\text{f}}Y\right] = -\text{Im}\left(\text{tr}\left[\rho_nU_n\right]\right) =- \frac{1}{2^{n+1}} \text{Im}\left(\text{tr}\left[U_n\right]\right).
\end{align}
In order to estimate the expected values of the operators in the equations labelled~\eqref{eq:XY_estimation}, one needs to evolve the quantum system repeatedly to collect enough statistics through measurements performed on the control qubit.

The number of measurements required to estimate the trace with an accuracy $\varepsilon$ is in $\mathcal{O}(\text{ln}\left(1/\delta\right)/\varepsilon^2$), where $\delta$ is the probability that the estimate
is farther from the true value than $\varepsilon$~\cite{KL98,DFC05}. Hence, the complexity of estimating the trace with a fixed accuracy is independent of the number of qubits and scales logarithmically with the error probability. This makes DQC1 an efficient method for estimating the normalized trace of unitary operators~\cite{DFC05}.

\subsection{Quantum Kernel Machine Learning Using DQC1}
One interesting application of DQC1 is estimating the trace of a unitary operator, which has applications in the development of classically intractable quantum kernel machines~\cite{GZO19}. In what follows, we explain the mathematical procedure for using DQC1 in quantum kernel machines.

To start, let us rewrite Eq.~\eqref{eq:kernel_element} for the case of a quantum kernel function as
\begin{equation}
    \label{eq:quantum_kernel_element}
    k(\bm{x}_i,\bm{x}_j):=\braket{\Phi(\bm{x}_i)}{\Phi(\bm{x}_j)}_{\mathcal{H}},
\end{equation}
where $\Phi(\cdot)$ represents a quantum feature mapping~\cite{GZO19} and $\braket{\cdot}{\cdot}$ denotes the overlap of two quantum states in the quantum feature space (i.e., Hilbert space) $\mathcal{H}$. 

In order to implement a quantum kernel based on Eq.~\eqref{eq:quantum_kernel_element} using DQC1, we first express $U_n$ (see Fig.~\ref{fig:DQC1}b), which acts on the register qubits, as a sequential multiplication of two unitary operators:
\begin{equation}
    \label{eq:register_q_decomposition}
    U_n(\bm{x}_i,\bm{x}_j)=U_D^\dagger(\bm{x}_i)U_D(\bm{x}_j),
\end{equation}
where each of the feature data $\bm{x}_i$ and $\bm{x}_j$ are encoded into the $U_n$ operator (see Fig.~\ref{fig:DQC1}b). As shown in Fig.~\ref{fig:DQC1}c, the operator $U_D(\bm{x}_i)$, $i \in \{1,\ldots, N\}$, can be expressed as the product of $D$ encoding-block operators $V_d(\bm{x}_i)$, $d\in\{1,\ldots,D\}$, and Hadamard gates $H$: 
\begin{equation}
    U_D(\bm{x}_i)=V_{D}(\bm{x}_i)H^{\otimes{n}}\cdots V_{1}(\bm{x}_i)H^{\otimes{n}}.
\end{equation}
The encoding-block operators $V_{d}(\bm{x}_i)$ can take different encoding patterns for each encoding block where the following encoding pattern is used to encode any feature data into the quantum circuit~\cite{HCT+19}:
\begin{equation}
    \label{eq:encoding}
    V_{d}(\bm{x}_i)=\text{exp}\left(\text{i}\sum_{C \in {\mathcal{S}}}g_C(\bm{x}_i)\prod_{k\in C}Z_k\right),
\end{equation}
where $Z$ refers to the Pauli-$Z$ operator, $\mathcal{S}$ refers to the set of all non-empty subsets of $\{ 1, 2,\ldots, n \}$, and $g_{C}(\cdot): \mathcal{X} \rightarrow{\mathbb{R}}$ is the encoding function. For instance, for the case of $\bm{x}_i \in{\mathbb{R}^{2}}$, $g_{\{u\}}(\bm{x}_i)=x_u$, $u\in\{1,2\}$, and $g_{\{1,2\}}(\bm{x}_i)=(\pi-x_1)(\pi-x_2)$~\cite{HCT+19}. 

Using Eqs.~~\eqref{eq:XY_estimation}--\eqref{eq:register_q_decomposition} and preparing the register qubits at a pure initial state $\rho_n=\dyad{\psi}{\psi}$, we can relate the trace estimation of DQC1 to the kernel function estimation as follows:
\begin{align}
    \label{eq:trace_to_kernel}
    \text{tr}\left(\rho_nU_n(\bm{x}_i,\bm{x}_j)\right) & =
    \text{tr}\left(\rho_nU^\dagger_D(\bm{x}_i)U_D(\bm{x}_j)\right)\\ \nonumber
    & = \text{tr}\left(\dyad{\psi}{\psi}U^\dagger_D(\bm{x}_i)U_D(\bm{x}_j)\right) \\ \nonumber
    & =  \braket{\Phi(\bm{x}_i)}{\Phi(\bm{x}_j)}_{\mathcal{H}}\\ \nonumber
    & =k(\bm{x}_i,\bm{x}_j),
\end{align}
where $\Phi(\bm{x}_j):=U_D(\bm{x}_j)\ket{\psi}$.

\subsection{Multiple Kernel Learning} \label{subsec: MKL}
While the representer theorem provides the mathematical framework for using kernel functions for function approximation, the performance of kernel-based methods depends on the ability of the kernel function to approximate the underlying distribution in the dataset. To this end, in addition to standard kernels such as linear, polynomial, and Gaussian, there are other types of customized kernels that are used for specific applications, such as natural language processing~\cite{FCC+15,HCJ+02}, pattern recognition~\cite{SC04}, and bioinformatics~\cite{NPC+16,NVC+13}.

In recent years, MKL has been used to boost the expressivity of kernel machines. The idea behind MKL is to use a combination of kernel functions instead of a single kernel function for learning. Kernel functions can be combined in many ways. In this work, we focus on two forms of kernel combinations. The first form is a linear combination of $M$ different kernels:
\begin{equation}
\label{eq:linear_sum_of_kernels}
    k(\bm{x}_i,\bm{x}_i) = \sum_{m=0}^M\alpha_m k_m(\bm{x}_i,\bm{x}_j,\bm{\theta}_m),
\end{equation}
where $\bm{\alpha} = [\alpha_m ]_{m=1,\ldots,M} \in \mathbb{R}^M$ is the vector of kernel weights and $\bm{\theta}_m\in\mathbb{R}^{q_m}$, $q_m\in\mathbb{N}$, parameterizes each kernel function. We call $\bm{\mathit{\Theta}}:=\{\bm{\theta}_1,\ldots,\bm{\theta}_M\}$ the set of all kernel parameters. In the second form, kernels are combined multiplicatively as follows:
\begin{equation}
    \label{eq:kernel_product}
    k(\bm{x}_i,\bm{x}_j) = \prod_{m=0}^M k_m(\bm{x}_i,\bm{x}_j,\bm{\theta}_m).
\end{equation}

There are various methods for choosing $\bm{\alpha}$ and $\bm{\mathit{\Theta}}$. These methods include heuristic~\cite{TBN+08,QL09}, optimization-based~\cite{LCB+04,KSC02,HCX08,OSW05}, Bayesian~\cite{GR05,DG09}, and boosting~\cite{BME02,CKS02} approaches. In practice, the performance of such methods depends on the structure of the dataset and the mathematical form of the combined kernel. In our simulations of the QMKL method, without loss of generality, we use an optimization-based method for choosing $\bm{\alpha}$ and $\bm{\mathit{\Theta}}$.

One important component in optimization-based methods is the objective function. Here, we choose the empirical risk functional, $R[f]$, as the objective function:
\begin{equation}
\label{eq:empirical_risk}
    R[f(\bm{x})] = \frac{1}{N} \sum_{i=1}^N L(f(\bm{x}_i), y_i)\,,
\end{equation}
where $f(\bm{x})$ is obtained through Eq.~\eqref{eq:rep_theorem} for the combined kernel, and $L$ refers to a loss function that measures the discrepancy between $f(\bm{x}_i)$ and the actual target value $y_i$. 

The optimal choice for the function $f$, denoted by $f^*$, is the one that minimizes the empirical risk:
\begin{equation}
\label{eq:Risk_minimization}
    \min_{\substack{\bm{\beta},\bm{\Gamma}}} \, R\left[f \right],
\end{equation}
where $\bm{\beta}$ is the vector of the base-learner parameters and $\bm{\mathit{\Gamma}}:=  {\{\bm{\alpha},\bm{\mathit{\Theta}}\}}$. One way to minimize the empirical risk is to alternate between optimizing over $\bm{\mathit{\Gamma}}$ and $\bm{\beta}$ during the minimization procedure. This is done by fixing $\bm{\mathit{\Gamma}}$ and minimizing the empirical risk with respect to $\bm{\beta}$ and vice versa. We follow this approach in our simulations.

\section{Quantum multiple kernel learning}
\label{sec:qmkl}
In this section, we first present different ways of combining kernels, including linear and multiplicative methods. We then outline the details of the optimization problems for each of the proposed multiple quantum kernels.

\subsection{Linear Kernel Combination}\label{subsec: linear combination}
To start, we present our QMKL method for the case of additive kernel combination. From Eq.~\eqref{eq:trace_to_kernel}, when the register qubits are initialized in a pure state, the resultant kernel function from the circuit matches the kernel definition, that is, $k(\bm{x}_i,\bm{x}_j) := \braket{\Phi(\bm{x}_i)}{\Phi(\bm{x}_j)}_{\mathcal{H}}$. On the other hand, when the initial state of the register qubits is a parameterized mixed state,
\begin{equation} \label{eq: parameterized state}
    \rho_n=\sum_{m=0}^{M=2^n -1}\alpha_m \dyad{m}{m},
\end{equation}
where the sum of the state's parameters (kernel weights) is one:
\begin{equation}
    \sum_{m=0}^{M=2^n-1} \alpha_m = 1.
\end{equation}
Now, rewriting Eq.~\eqref{eq:trace_to_kernel} results in the following:
\begin{align} \label{eq:trace_to_kernel_fully_state} \nonumber
    \text{tr}\left(\rho_nU_n(\bm{x}_i,\bm{x}_j)\right) & =
    \text{tr}\left(\sum_{m=0}^{M}\alpha_m\ketbra{m}{m} U_D^\dagger(\bm{x}_i)U_D(\bm{x}_j) \right)\\ \nonumber
    & =  \sum_{m=0}^{M}\alpha_m \text{tr}\left(\bra{m} U_D^\dagger(\bm{x}_i)U_D(\bm{x}_j) \ket{m} \right)\\ \nonumber
    & = \sum_{m=0}^{M} \alpha_m \braket{\Phi_m(\bm{x}_i)}{\Phi_m(\bm{x}_j)} \\
    & = \sum_{m=0}^{M}\alpha_m k_m(\bm{x}_i,\bm{x}_j),
\end{align}
where~$\Phi_m(\bm{x}_j):=U_D(\bm{x}_j)\ket{m}$. In other words, by initializing the register qubits in a parameterized mixed state, the trace is a combined quantum kernel with each $\alpha_m$ as kernel weights.

Note that we can also develop a linear combination of parameterized quantum kernels. This type of kernel is a special form of the kernel that we discuss in Section~\ref{subsec: additive multiplicative}.
\begin{figure*}[ht]
    \centering
    \includegraphics[width=0.6\textwidth]{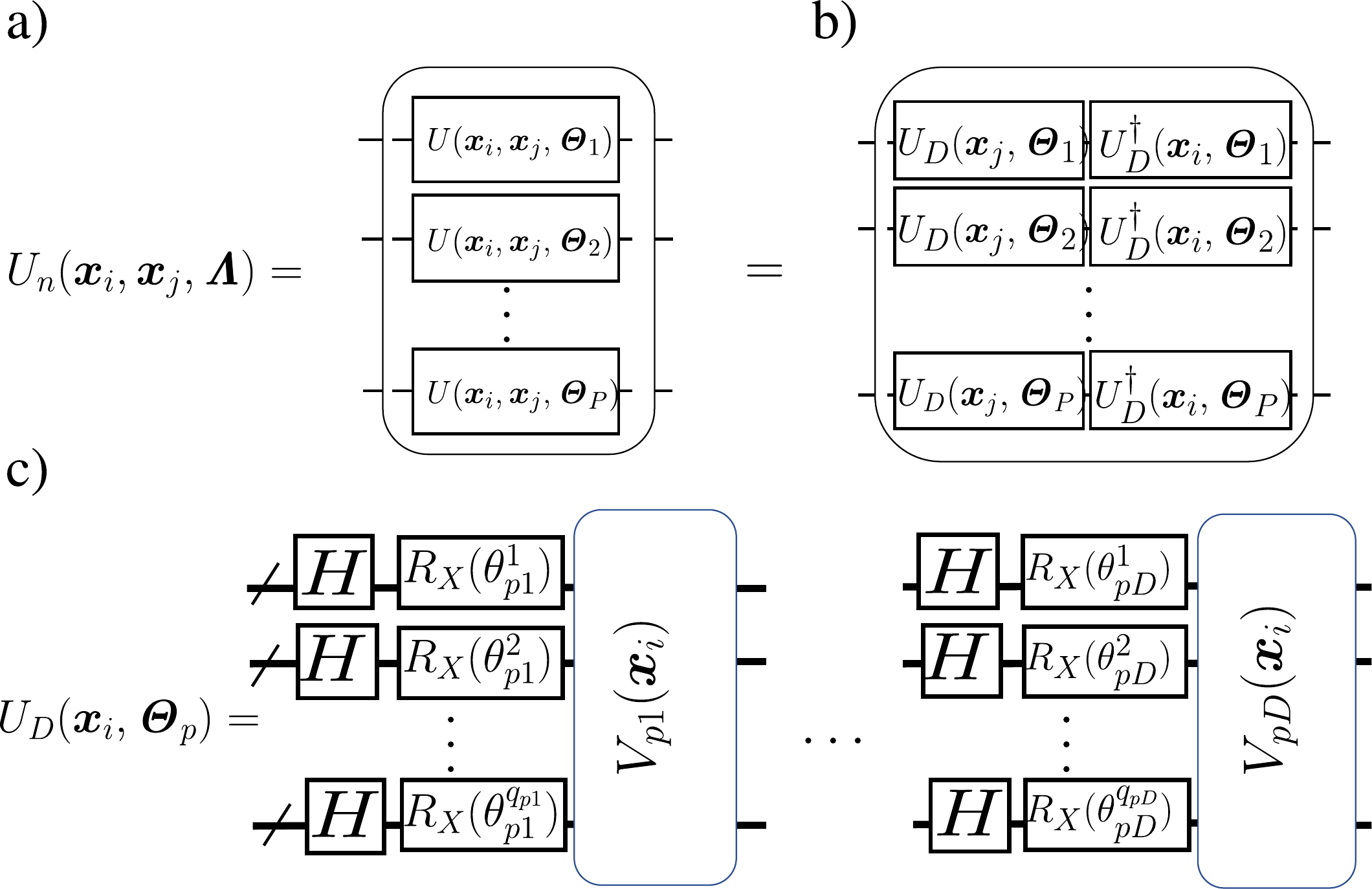}
    \caption{Schematic of a quantum circuit for generating a product of $P$ different kernels. a) The $U_n(\bm{x}_i,\bm{x}_j,\bm{\mathit{\Lambda}})$ operator (see Fig.~\ref{fig:DQC1}a) expressed as a tensor product of $P$ unitary operators. b) Each $U(\bm{x}_i,\bm{x}_j,\bm{\mathit{\Theta}}_p)$ is decomposed into a product of two unitary operators, where each contains information about a data sample and kernel parameters $\bm{\mathit{\Theta}}_p$, $p\in\{1,\ldots,P\}$. c) One possible way for encoding $\bm{x}_i$ and $\bm{\mathit{\Theta}}_p$ by repeating the encoding block $D$ times. For simplicity, we consider the same encoding block as Eq.~\eqref{eq:encoding}. In this work, we consistently use $p$ and $d$ as iterators that refer to the $p$-th register and $d$-th encoding block, respectively.}
    \label{fig:QMKL}
\end{figure*}
 
\subsection{Multiplicative Kernel Combination}
In this section, we discuss how a quantum kernel defined as a multiplicative combination of different kernels can be constructed using DQC1. Figure~\ref{fig:QMKL} shows the corresponding quantum circuit for constructing such a combined kernel. First, $n$ register qubits are partitioned into $P$ subsets, denoted by $s_{p}$, $p \in \{1, 2,\ldots, P \}$. The number of qubits in each subset is the subset's cardinality~$\vert s_p \vert$. Let us define $\bm{\mathit{\Theta}}_p:=\{\bm{\theta}_{p1},\cdots,\bm{\theta}_{pD}\}$ as the kernel parameters associated with the $p$-th subset, where $\bm{\theta}_{pd}$, $d \in \{1, 2,\ldots, D \}$, is a $q_{pd}$-dimensional vector, $q_{pd} \in \mathbb{N}$. We also define $\bm{\mathit{\Lambda}}$ as the set of kernel parameters for all $P$ registers, $\bm{\mathit{\Lambda}} := \{{\bm{\mathit{\Theta}}_1},\cdots,{\bm{\mathit{\Theta}}_P}\}$. Then, as shown in Fig.~\ref{fig:QMKL}a, we can decompose $U_n$ into a tensor product of $P$ parameterized unitary operators:
  
\begin{align}
    \label{eq:Un_tensor_prod_decomposition}
    & U_n(\bm{x}_i,\bm{x}_j,\bm{\mathit{\Lambda}}) = \\ \nonumber
    &U(\bm{x}_i,\bm{x}_j,{\bm{\mathit{\Theta}}}_P) \!\otimes{ \ U(\bm{x}_i,\bm{x}_j,\bm{\mathit{\Theta}}_{P-1})} \! \otimes{ \! \cdots} \!\otimes{ \!U(\bm{x}_i,\bm{x}_j,\bm{\mathit{\Theta}}_{1})},
\end{align}
where each $U(\bm{x}_i,\bm{x}_j,\bm{\mathit{\Theta}}_p)$ acts on the $p$-th register subset. Following the same approach as we did for Eq.~\eqref{eq:register_q_decomposition}, we express each of the $U(\bm{x}_i,\bm{x}_j,\bm{\mathit{\Theta}}_p)$ as the product of two unitary operators (see Fig.~\ref{fig:QMKL}b):
\begin{equation}
    \label{eq:sub_register_q_decomposition}
    U(\bm{x}_i,\bm{x}_j,\bm{\mathit{\Theta}}_p) := {U^\dagger_D}(\bm{x}_i,\bm{\mathit{\Theta}}_p){U_D}(\bm{x}_j,\bm{\mathit{\Theta}}_p),
\end{equation}
where $U_D(\bm{x}_i,\bm{\mathit{\Theta}}_p)$ contains the information about data sample $\bm{x}_i$ and kernel parameters $\bm{\mathit{\Theta}}_p$. The choice of $U_D(\bm{x}_i,\bm{\mathit{\Theta}}_p)$ can be any unitary operator which is inefficient to simulate classically; here, we choose~\cite{HCT+19} (see Fig.~\ref{fig:QMKL}c):
\begin{equation}
    \label{eq:multiple_kernel_encoding_parameterization}
    U_D(\bm{x}_j,\bm{\mathit{\Theta}}_p) =\prod_{d=1}^D{V_{pd}(\bm{x}_j){U(\bm{\theta}_{pd})}{H^{\otimes{\vert s_p \vert}}}},
\end{equation}
where one choice~\cite{HCT+19} for $U(\bm{\theta}_{pd})$ is:
\begin{equation}
U({\bm{\theta}}_{pd})= \otimes_{k=1}^{\vert q_{pd} \vert}{\text{e}^{\mathrm{i}\theta_{pd}^k{W}}},
\end{equation}
where $W$ is one of the Pauli-$\{X, Y, Z\}$ operators~\cite{NC02}.

We are now ready to derive a multiplicative kernel combination using the circuit in Fig.~\ref{fig:QMKL}. Let us assume that the register qubits are initialized in a pure state $\rho_n=\rho_1\otimes\cdots\otimes{\rho_P}$, where $\rho_p=\ketbra{\psi_p}{\psi_p}$, $p\in\{1,\ldots,P\}$. Following a similar approach as in Eq.~\eqref{eq:trace_to_kernel} and using~\eqref{eq:Un_tensor_prod_decomposition}--\eqref{eq:sub_register_q_decomposition}, we have
\begin{equation}
\begin{split}
    \label{eq:trace_to_product_kernel}
    \text{tr}\left(\rho_nU_n(\bm{x}_i,\bm{x}_j,\bm{\mathit{\Lambda}})\right) & = \text{tr}\left(\bigotimes_{p=1}^P{\rho_p{U^\dagger_D}(\bm{x}_i,\bm{\mathit{\Theta}}_p)U_D(\bm{x}_j,\bm{\mathit{\Theta}}_p)}\right) \\
    &=\prod_{p=1}^P\text{tr}\left({\ketbra{\psi_p}{\psi_p}{{U^\dagger_D}}(\bm{x}_i,\bm{\mathit{\Theta}}_p)U_D(\bm{x}_i,\bm{\mathit{\Theta}}_p)}\right)\\
    &= k(\bm{x}_i,\bm{x}_j,\bm{\mathit{\Theta}}_1)\cross\cdots\cross{k(\bm{x}_i,\bm{x}_j,\bm{\mathit{\Theta}}_P)} \\
    & = k(\bm{x}_i,\bm{x}_j,\bm{\mathit{\Lambda}}).
\end{split}
\end{equation}
Thus, $\text{tr}\left(\rho_nU_n(\bm{x}_i,\bm{x}_j,\bm{\mathit{\Lambda}}) \right)$ is equivalent to a product of $P$ parameterized kernels. We can then use DQC1 to estimate the trace of $\rho_nU_n(\bm{x}_i,\bm{x}_j,\bm{\mathit{\Lambda}})$, thereby evaluating the combined kernel $k(\bm{x}_i,\bm{x}_j,\bm{\mathit{\Lambda}})$.

\subsection{Additive Multiplicative Kernel Combination}\label{subsec: additive multiplicative}
To derive Eq.~\eqref{eq:trace_to_product_kernel}, we assume that each $\rho_p$ is initialized in a pure state. We now consider an \mbox{$M_p$-dimensional} vector $\bm{\alpha}_p$, $\bm{\alpha}_p = [\alpha_{pi}] \in \mathbb{R}^{2^{\vert s_p \vert}}$, to initialize the $p$-th register in a 
parameterized mixed state \mbox{$\rho_p=\sum_i\alpha_{pi}\ketbra{i}{i}$, $i \in \{0,\cdots, M_p=2^{\left|s_p\right|} -1\}$}. Defining $\bm{A}:=\{\bm{\alpha}_1,\ldots,\bm{\alpha}_p,\ldots,\bm{\alpha}_P\}$ as the set of all initial states' parameters, it is straightforward to derive the additive multiplicative kernel combination for the case of paramterized mixed states as follows:
\begin{equation}
\begin{split}
    \label{eq:sum_prod_kernel}
    & \text{tr}\left(\rho_n(\bm{A})U_n(\bm{x}_i,\bm{x}_j,\bm{\mathit{\Lambda}})\right) =  \\
    & \sum_l\alpha_{1l}k_l(\bm{x}_i,\bm{x}_j,\bm{\mathit{\Theta}}_1)\cross\cdots\cross{\sum_l\alpha_{Pl}k_l(\bm{x}_i,\bm{x}_j,\bm{\mathit{\Theta}}_P)}.
\end{split}
\end{equation}

The kernel in Eq.~\eqref{eq:sum_prod_kernel} reduces to a linear sum of parameterized kernels for the case of $P=1$, where $U\left(\bm{x}_i,\bm{x}_j,\bm{\mathit{\Theta}}_1\right)$ acts on the entire $n$-qubit register. The resulting kernel combination becomes the linear kernel combination discussed in Section~\ref{subsec: linear combination}, where each individual kernel is parameterized by $\bm{\mathit{\Theta}}_1$.

\subsection{Optimization Problem}
As discussed in Section~\ref{subsec: MKL}, an optimization problem can be solved to find the optimal parameters of the combined kernels as well as the kernel weights. Considering the empirical risk represented by Eq.~\eqref{eq:empirical_risk} as the objective function of the optimization problem, for the case of a linear kernel combination, the optimization problem can be formulated as follows:
\begin{align}
    \label{eq:constraint_linear_kernel}
    \min_{\bm{\alpha}} \, &R\left[f \right], &\\ \nonumber
    &\sum_{m =0}^M{\alpha}_m=1, & \\ \nonumber
    & \, 0 \leq \alpha_m,  & \forall \, m = 1,\ldots, M.
\end{align}
The above constraints follow the properties of the density matrix, where it is a positive semi-definite operator with its trace equal to one.

For a general parameterized mixed state, the number of kernel combination parameters (where a kernel parameter is denoted $\alpha_{m}$), grows exponentially with the number of register qubits because $M = 2^n-1$. As $M$ grows exponentially with $n$, having such a large number of parameters makes optimizing the kernel combination an intractable problem. To avoid such a scenario, we consider a \emph{restricted} set of mixed states where the number of the state's parameters, $M$, grows linearly with the number of register qubits. Note that even for restricted mixed states, the number of individual kernels could still grow exponentially with the number of register qubits.
It is worth mentioning that,
unlike in the case of classical MKL methods where one must calculate each of the kernel functions individually,
here we use DQC1 to estimate the combined kernel function without needing to calculate each of the individual kernels explicitly.

For the case of multiplicative and additive multiplicative kernels, the optimization problem takes the form
\begin{align}
\label{eq:Risk_minimization_multiplicative}
    \min_{\substack{\bm{\beta},\bm{\mathit{\Omega}}}} \, & R\left[f \right] & \\ \nonumber
    &\Vert \bm{\alpha}_p \Vert_1=1, & \forall \, p \in \{1,\ldots, P \},\\ \nonumber
    & \, 0 \leq  \alpha_{pi}, & \forall \, p \in \{1,\ldots, P \},\forall \, i\in\{1,\cdots, 2^{\left|s_p\right|}\},
\end{align}
where $\bm{\mathit{\Omega}}:=\{\bm{A},\bm{\mathit{\Lambda}}\}$ and $\Vert \cdot \Vert_1$ is the $L_1$-norm. Once again, the constraints are imposed by the properties of the density matrix. Depending on the type of kernel, a certain subset $\bm{\mathit{\Omega}}$ is chosen for optimization.

As with the linear kernel combination, we can restrict the number of state parameters to simplify the optimization problem as well as the initial state preparation, that is, $\bm{\alpha}_p\in{\mathbb{R}^{M_p}}$ such that $M_p\in{\mathbb{N}}$ grows linearly with $\vert s_p \vert$.

\section{Simulation settings}
\label{sec:simulation}
To assess the performance of one type of a kernel over others for a given machine learning task, one must perform a benchmarking procedure. Given the size of, and level of noise present in, available quantum computers, we classically estimate the quantum kernel to demonstrate the performance of QMKL on classification problems. This classical simulation of the quantum system, however, becomes computationally challenging as the number of qubits increases.
For this reason, we limit our simulations to quantum systems with fewer than five qubits. We choose to simulate a linear combination of quantum kernels as represented in Eq.~\eqref{eq:trace_to_kernel_fully_state} as a proof of concept because this choice is largely supported by the success of MKL algorithms that employ a linear sum of kernels in a variety of applications~\cite{CTS+17,TSM+19}.

We consider two binary classification tasks to investigate the performance of our QMKL method. We perform each classification task with three different quantum kernel learning models. First, we consider a single quantum kernel learning (SQKL) model by initializing the register qubits in a pure state, that is, $\rho_n=(\ketbra{0})^{\otimes{n}}$. 
Second, we initialize the register qubits in a fully mixed state and call the resultant model ``fixed-QMKL'', as the kernel weights are equal. For the last model, we parameterize the initial state of the register qubits according to Eq.~\eqref{eq: parameterized state} and choose the parameters of the model to minimize the empirical risk represented by Eq.~\eqref{eq:empirical_risk} where we consider a 0--1 loss function~\cite{SB14}. We refer to this last setting simply as ``QMKL''. For each of the three models, we feed the estimated kernel matrix from the quantum kernel into a support vector machine to perform the classification task.

For the case of the QMKL model, we solve the optimization problem formulated as~\eqref{eq:constraint_linear_kernel}. As explained in Section~\ref{subsec: MKL}, the optimizer alternates between optimizing over $\bm{\alpha}$ and $\bm{\beta}$ during the minimization procedure in order to find the best model parameters.

\begin{figure}[ht]
    \centering
    \includegraphics[width=\columnwidth]{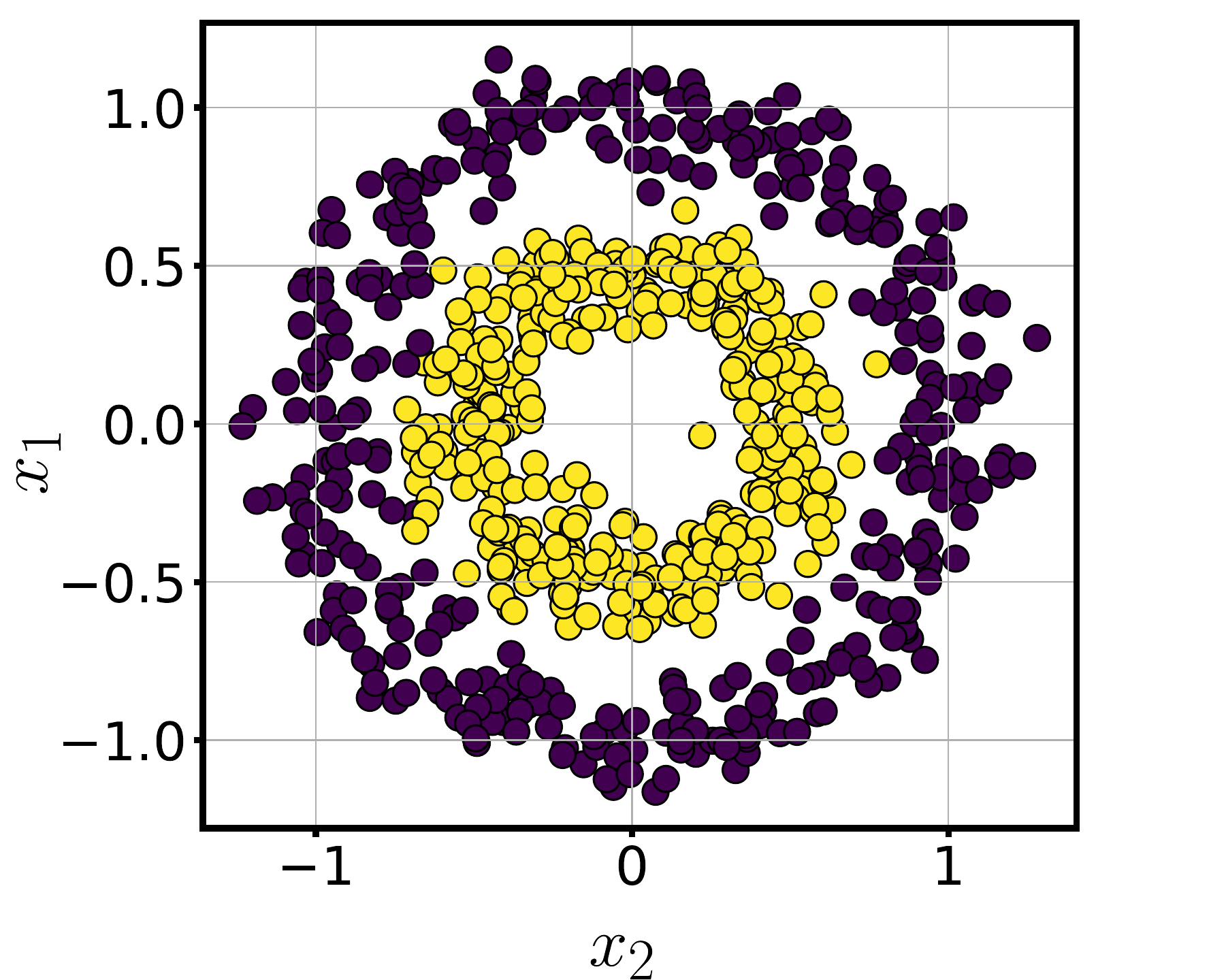}
    \caption{Representation of the “circles” dataset generated using the \mbox{{\tt datasets}} module from the \mbox{{\tt scikit-learn}} package in Python. Each class (yellow or purple) consists of 350 data samples. The ratio of the inner circle's \mbox{(class 1)} radius to that of the outer circle \mbox{(class 2)} is $0.8$, and the standard deviation of the Gaussian noise added to the data is $0.1$.}
    \label{fig:Circle_dataset}
\end{figure}

 The first classification problem includes a two-dimensional synthetic ``circles'' dataset (see Fig.~\ref{fig:Circle_dataset}). For this problem, the corresponding hyperparameters of the three aforementioned models are the repetition number of the encoding blocks $d$ and the optimization algorithm's parameters for the case of QMKL (see Appendix~\ref{app:circle dataset}). To perform hyperparameter tuning over the synthetic dataset, we generate $20$ random instances of this dataset (see one representation in Fig.~\ref{fig:Circle_dataset}). We then randomly split the data samples into training and test datasets, with 75\% of the data used for the training dataset and the rest for the test dataset. To restrict the complexity of the optimization in QMKL, we use only 50\% of the training dataset. Using the best set of parameters, we then re-train each model on another set of 100 randomly generated instances of the circles dataset with the same training and test split ratio while using only 50\% of the training data for optimization.
 
 We compute the average $\mu$ and the standard deviation $\sigma$ of the classification accuracy of the trained models over these 100 datasets. The classification accuracy is defined as the percentage of the correctly classified data samples with respect to the total number of unseen data samples. Note that $\sigma$ reflects the standard deviation of the model's performance on different datasets, and not the model's error itself.
 
For the second classification problem, we consider the ``German Credit Data'' dataset from~\cite{DC17}. Whereas the dataset consists of 1000 data samples, each with 20 features, we only use four of the features (see Appendix~\ref{app:german_dataset}) for training the models. Using these four features, the classification accuracy is only slightly worse than when all 20 features are used~\cite{Zur10}. Keeping the training and test split ratios identical to those for the  synthetic dataset, we use different random seeds for splitting the dataset into 20 instances of training and test datasets. Again, only 50\% of the training data is used in the QMKL optimization step. These 20 different splits are used to tune the hyperparameters of the models. Finally, we compute the average and standard deviation of the classification accuracy of the trained models over 100 different training and test splits of the dataset for the best set of parameters.

\section{Results}
\label{sec:results}
In this section, we report on the performance of the three quantum kernel models, namely, SQKL, fixed-QMKL, and QMKL, for the two classification tasks. 

\begin{figure}[ht]
    \centering
    \includegraphics[width=\columnwidth]{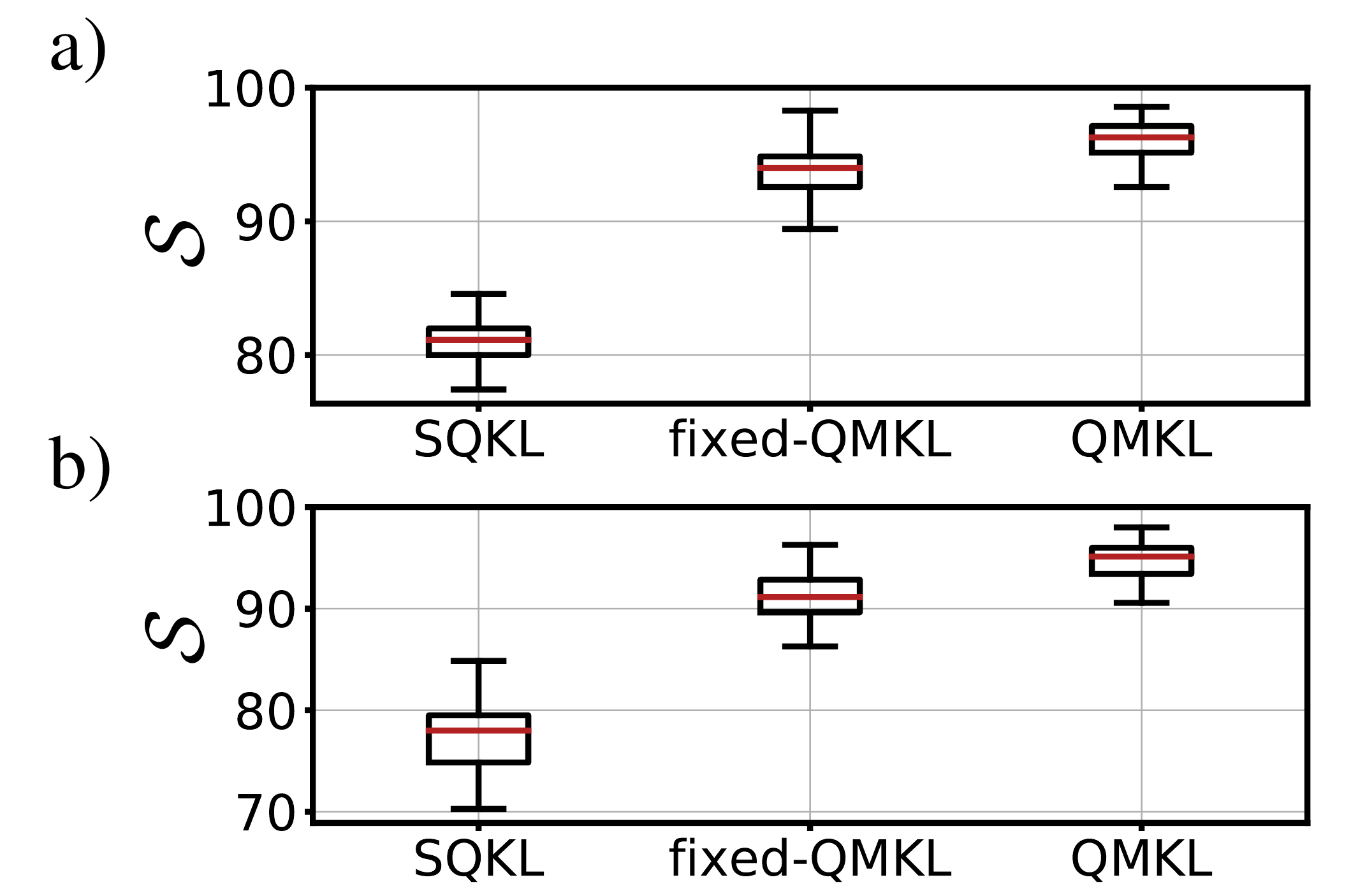}
    \caption{Statistics of the classification accuracy, $\mathcal{S}$, of the three quantum kernel methods on 100 instances of the synthetic circles dataset (see Fig.~\ref{fig:Circle_dataset}). The performance of the models on the a) training and b) test datasets is shown.}
    \label{fig:circle_data_box_plot}
\end{figure}

We begin with the synthetic circles dataset. Table~\ref{tab:circle_data} summarizes the results of the three quantum kernel models on the synthetic dataset. We report the results for both training and test datasets.
\begin{table}[ht]
\centering
\begin{ruledtabular}
\begin{tabular}{lclcc}
 & \multicolumn{2}{c}{Training} & \multicolumn{2}{c}{Test} \\ \hline
Method & $\mu$ & \multicolumn{1}{c}{$\sigma$} & $\mu$ & $\sigma$ \\ \hline
SQKL & $81.17$ & $1.8$ & $77.53$ & $3.47$ \\ \hline
fixed-QMKL & $93.72$ & $1.87$ & $91.57$ & $2.9$ \\ \hline
QMKL & $96.05$ & $1.45$ & $94.52$ & $2.15$ \\
\end{tabular}
\end{ruledtabular}
\caption{Average, $\mu$, and  standard deviation, $\sigma$, of  classification accuracy of the three quantum kernels on 100 different instances of the circles dataset (see Fig.~\ref{fig:Circle_dataset}).}
\label{tab:circle_data}
\end{table}
Figure~\ref{fig:circle_data_box_plot} provides more statistical information about the performance of each of the quantum kernels on the 100 random realizations of the circles dataset. The vertical axis represents the classification accuracy, denoted by $\mathcal{S}$, of the models. The red line in each box shows the median classification score for the associated quantum kernel model. The lower and upper edges of the box respectively represent the 25th and 75th percentiles of $\mathcal{S}$. The lower whisker points to the minimum classification accuracy and the upper whisker points to the maximum classification accuracy achieved by each quantum kernel method.

\begin{figure}[ht]
    \centering
    \includegraphics[width=\columnwidth]{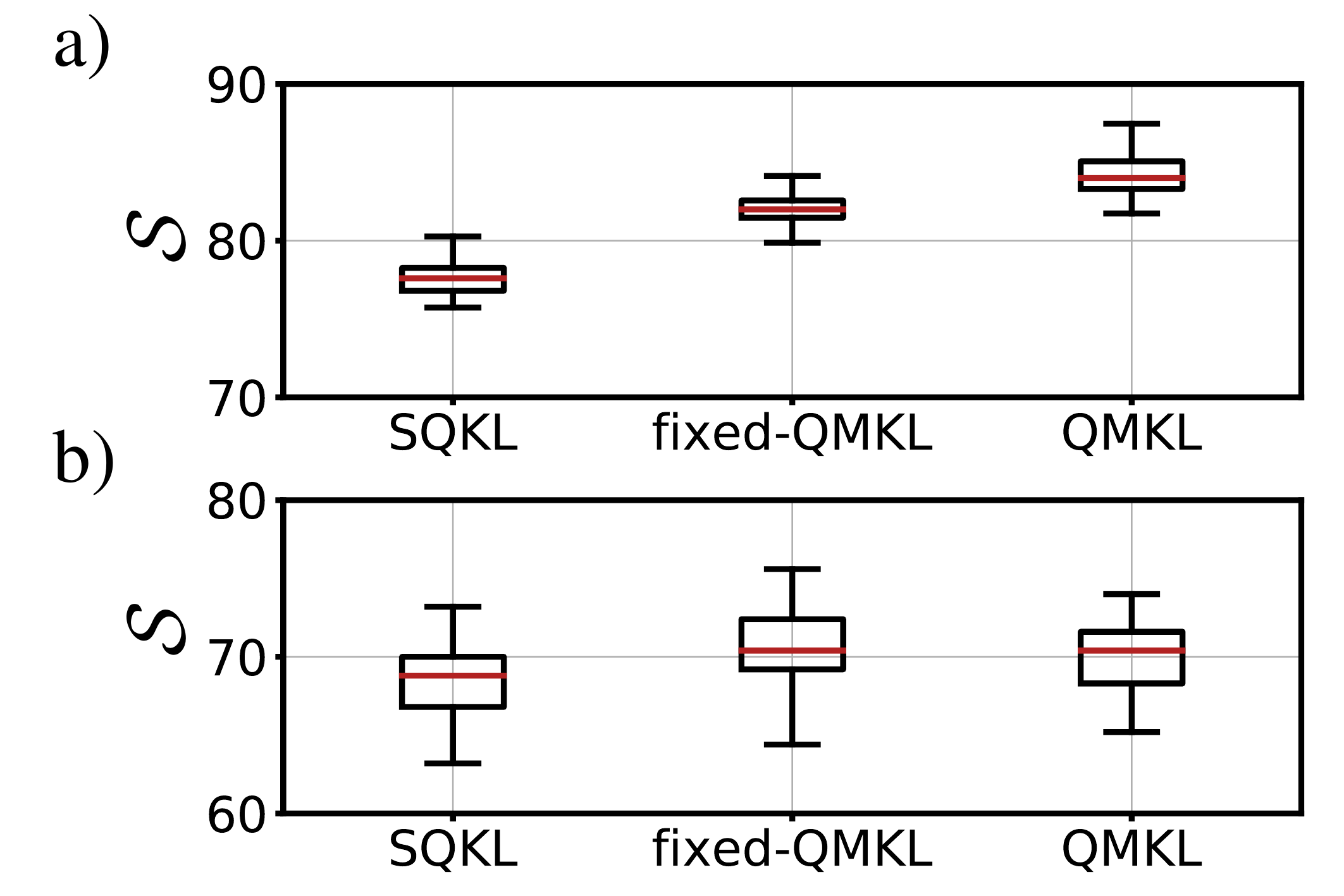}
    \caption{Statistics of the classification accuracy, $\mathcal{S}$, of the three quantum kernel methods on the 100 different splits of the German credit dataset. The performance of the models on the a) training and b) test datasets is shown.}
    \label{fig:german_data_box_plot}
\end{figure}

Table~\ref{tab:German_data} summarizes the results of the three trained models on the German credit dataset. The results for both the training and test datasets are reported. Further statistical analysis we performed on the accuracy of the trained models on this dataset are shown in the form of box plots in~Fig.~\ref{fig:german_data_box_plot}.

\begin{table}[ht]
\centering
\begin{ruledtabular}
\begin{tabular}{lclcc}
 & \multicolumn{2}{c}{Training} & \multicolumn{2}{c}{Test} \\ \hline
Method & $\mu$ & \multicolumn{1}{c}{$\sigma$} & $\mu$ & $\sigma$ \\ \hline
SQKL & $77.59$ & $1.00$ & $68.45$ & $2.22$ \\ \hline
fixed-QMKL & $82.02$ & $0.88$ & $70.38$ & $2.4$ \\ \hline
QMKL & $84.29$ & $1.49$ & $69.90$ & $2.12$ \\
\end{tabular}
\end{ruledtabular}
\caption{Average, $\mu$, and  standard deviation, $\sigma$, of  classification accuracy of the three different quantum kernels used on the German credit dataset.}
\label{tab:German_data}
\end{table}

\section{Discussion}
\label{sec:discussion}
In this section, we analyze the results of our proposed method for solving two classification problems. We wish to highlight that the current state of quantum hardware technologies and the computationally expensive simulation of quantum evolution prevent us from presenting a fair comparison of our method against current state-of-the-art classical machine learning algorithms. Therefore, in this work, we focus primarily on the improvement achieved through using multiple kernels in QMKL over SQKL.

For the synthetic circles dataset, the fixed-QMKL and QMKL methods improve the average test accuracy by $18.10\%$ and $21.69\%$, respectively, over the SQKL model. This observation shows the improvement achieved by using multiple kernels for classification for the synthetic dataset.

The QMKL model achieves the highest average accuracy on the 100 training datasets. This indicates that QMKL has a higher expressive power than the other two models via its combined kernel parameters. For all three models, a higher accuracy in the training phase correlates positively with the test accuracy. This indicates that a more complex kernel might be able to achieve even better test accuracy. Searching for a circuit architecture that achieves a more expressive kernel is outside the scope of this work, and thus is not discussed here.

Figure~\ref{fig:circle_data_box_plot} indicates that the QMKL model has greater classification accuracy quartiles compared to SQKL and fixed-QMKL models. This is also the case for the minimum and the maximum accuracy over 100 instances of the synthetic dataset. Consistently, both quantum kernel models that use multiple kernels outperform the SQKL model.

For the German credit dataset, similar to the circles dataset, both fixed-QMKL and QMKL models outperform the SQKL model by $2.81\%$ and $2.11\%$, where the improvement achieved by \mbox{fixed-QMKL} is slightly higher than the QMKL. As expected, a parameterized kernel, that is, QMKL, achieves a higher training accuracy than the other two models, as shown in Fig.~\ref{fig:german_data_box_plot}. However, QMKL  overfits on the training dataset, which results in a lower average accuracy over the test datasets. This is consistent with published results, in which the performance of neural networks is worse than that of a support vector machine with a linear kernel~\cite{Zur10}.

In Fig.~\ref{fig:german_data_box_plot}, the median classification accuracy of \mbox{fixed-QMKL} and QMKL is higher than that obtained from SQKL for both the training and test datasets. The minimum training accuracy of QMKL is greater than that achieved by almost half of the runs for the fixed-QMKL method. In general, the statistics shown by the box plots indicate that QMKL demonstrates more expressive power than both SQKL and fixed-QMKL. However, when it comes to the test accuracy, as already mentioned, overfitting on the training dataset reduces the generalization power, that is, the ability to predict the label of unseen data, of the QMKL model.

\section{Conclusion}
\label{sec:conclusion}
We have introduced a  multiple kernel machine learning method based on DQC1, which combines classically intractable kernels for machine learning applications. In particular, we have described the quantum circuit architectures for a linear sum, as well as a product, of individual kernels. We have solved two binary classification problems, including a synthetic dataset and a German credit dataset, as a proof of concept of our method. 

Using a linear combination of individual kernels, the method we have proposed has significantly improved the classification accuracy over that obtained from a single quantum kernel method. Using our approach, one can combine quantum kernels corresponding to different quantum computer architectures, thereby leveraging the power of all the available quantum resource simultaneously. We leave the study of the effect of different quantum circuits on the expressivity of the combined quantum kernel for future work.

\section{Acknowledgement} 
Barry C. Sanders acknowledges NSERC support.  Partial  funding  for  this  work  was  provided  by the  Mitacs Accelerate program.  We thank Marko Bucyk for reviewing and editing the manuscript.
We are grateful for the support we received from Maliheh Aramon, Gili Rosenberg, and  Elisabetta Valiante pertaining to the use of HOPE.

\appendix
\section{Hyperparameter Tuning}
In this section, we provide the details of the hyperparameter tuning method we use for both the synthetic and German credit datasets. 

\subsection{Circles Dataset}
\label{app:circle dataset}
\begin{figure}[ht]
    \centering
    \includegraphics[width=0.4\textwidth]{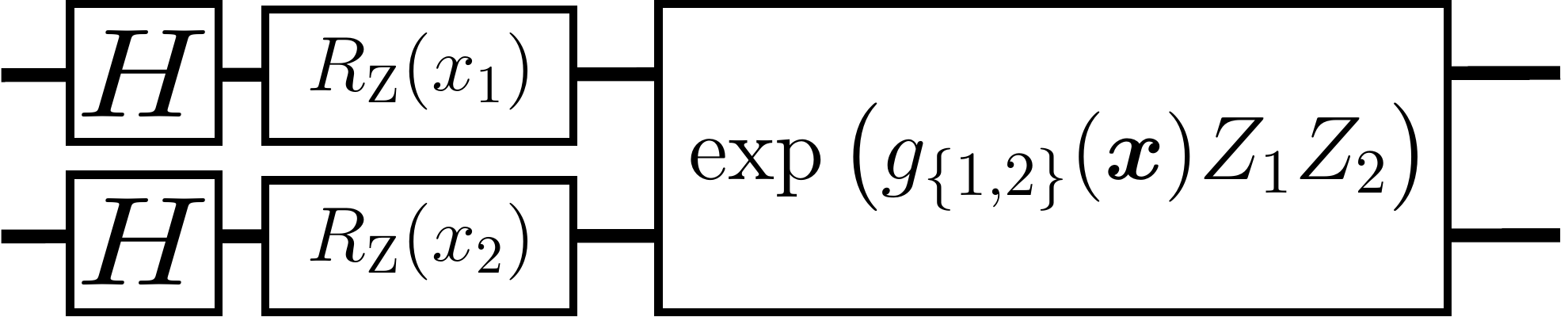}
    \caption{Encoding block used for the circles synthetic dataset.}
    \label{fig:circle_dataset_encoding}
\end{figure}

To train the quantum kernel models on the synthetic circles dataset,
we choose the encoding block in Fig.~\ref{fig:circle_dataset_encoding}. For QMKL, we search over the number of encoding block repetitions, $d$, as this parameter plays an important role in the complexity of the kernel. We use the selected value of $d$ through this search for both SQKL and fixed-QMKL. To optimize the kernel weights for the case of QMKL, we use the COBYLA optimization algorithm from Python's \mbox{{\tt scipy}} package. We realized that this optimizer is sensitive to its parameter ``rhobeg'', denoted by $h$ here; therefore, we  consider it in our hyperparameter tuning procedure. 
For hyperparameter tuning, we set the maximum number of iterations in COBYLA to 500 while default values are used for other arguments. For the case of QMKL, we define a parameter $r$ which represents the ratio of the training dataset considered in the optimization of the initial state parameters. While it is trivial that a greater value of $r$ is desired, to reduce the complexity of optimization, we explore how different values of $r$ affect the classification accuracy.

\begin{figure}[ht]
    \centering
    \includegraphics[width=0.49\textwidth]{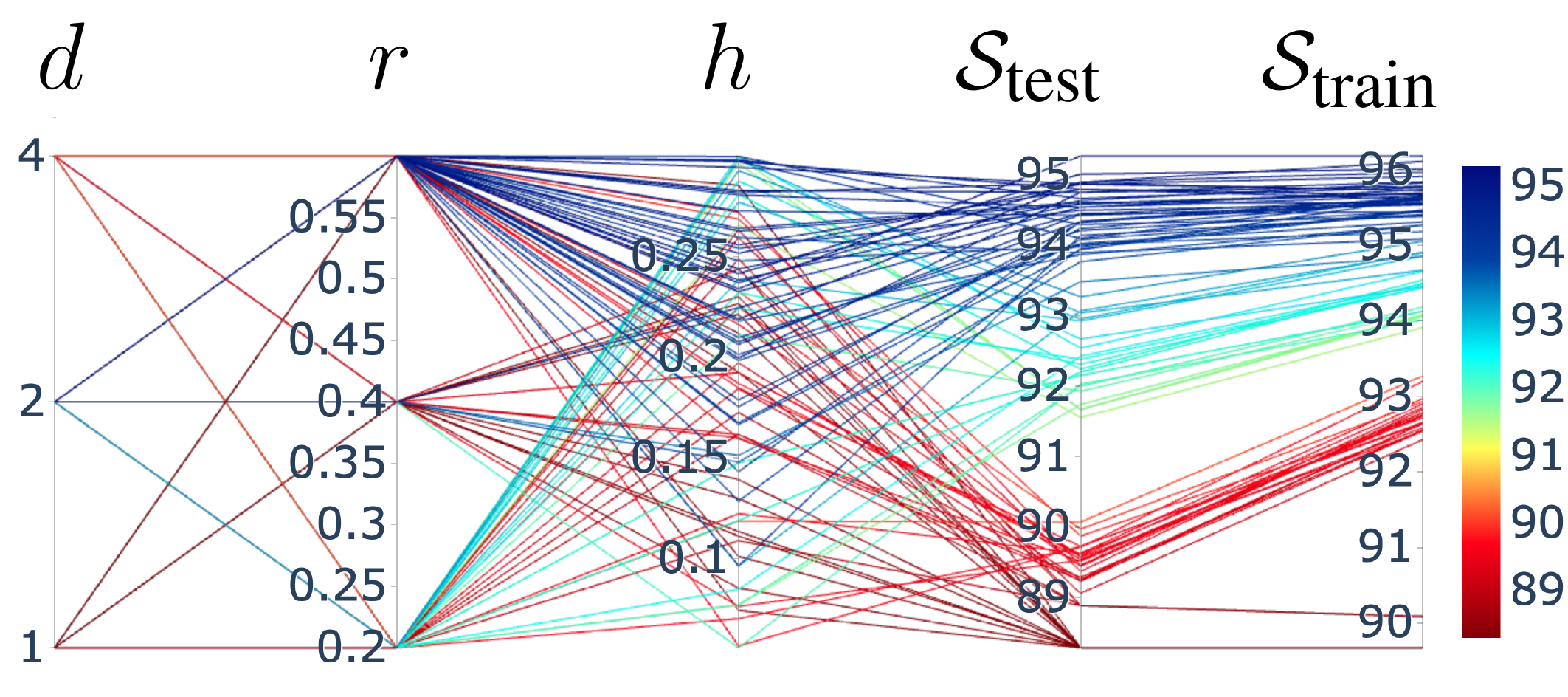}
    \caption{Results of hyperparameter tuning for the synthetic circles dataset.}
    \label{fig:hope_circle_dataset}
\end{figure}

For hyperparameter tuning, we use 1QBit's ``Hyperparameter OPtimization Environment'' (HOPE) package for automated tuning and benchmarking.
Figure~\ref{fig:hope_circle_dataset} shows the results of hyperparameter tuning on two parameters $d$ and $h$ when QMKL is used. The resultant best parameters are $d=2$ and $h=0.3$. We use these parameters and $r = 0.6$ to train QMKL (see Table~\ref{tab:circle_data}). For SQKL and fixed-QMKL, $r$ and $h$ are irrelevant. We use $d=2$ for these two models as well in order to have a fair comparison in terms of quantum resource usage for the three models.

\subsection{German Credit Dataset}
\label{app:german_dataset}
\begin{figure}[ht]
    \centering
    \includegraphics[width=0.485\textwidth]{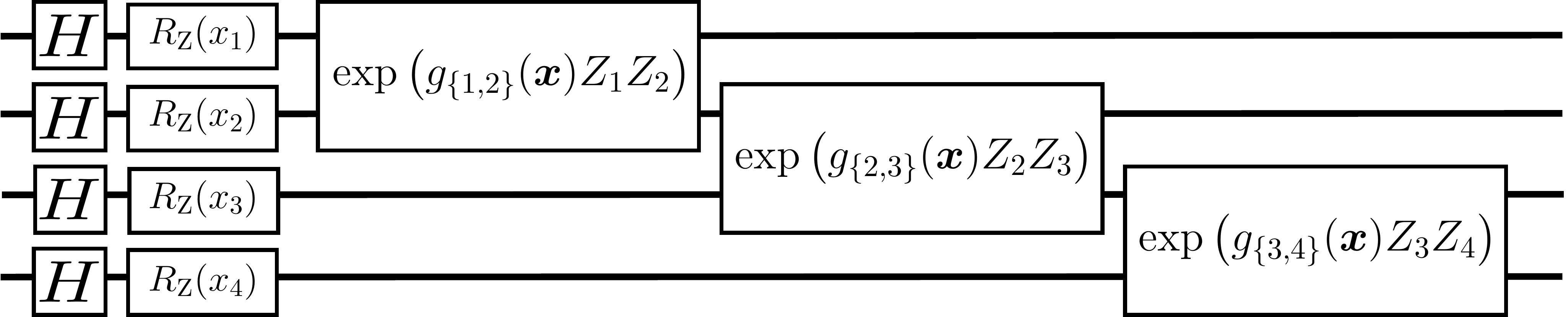}
    \caption{Encoding block used for the German credit dataset.}
    \label{fig:german_dataset_encoding}
\end{figure}

The German credit dataset contains 1000 data samples, each with 20 features. In our simulations, we choose only four out of 20 features, namely, ``chequing account existence'', ``duration'', ``credit history'', and ``employed since''. We use COBYLA with the same default setting as the one we use for the case of the circles dataset and employ the encoding block represented in Fig.~\ref{fig:german_dataset_encoding}. It is computationally expensive to simulate models using $r=1$; we therefore use an educated guess for the value of $r=0.5$ based on the results from the circles dataset.

\begin{figure}[ht]
    \centering
    \includegraphics[width=0.49\textwidth]{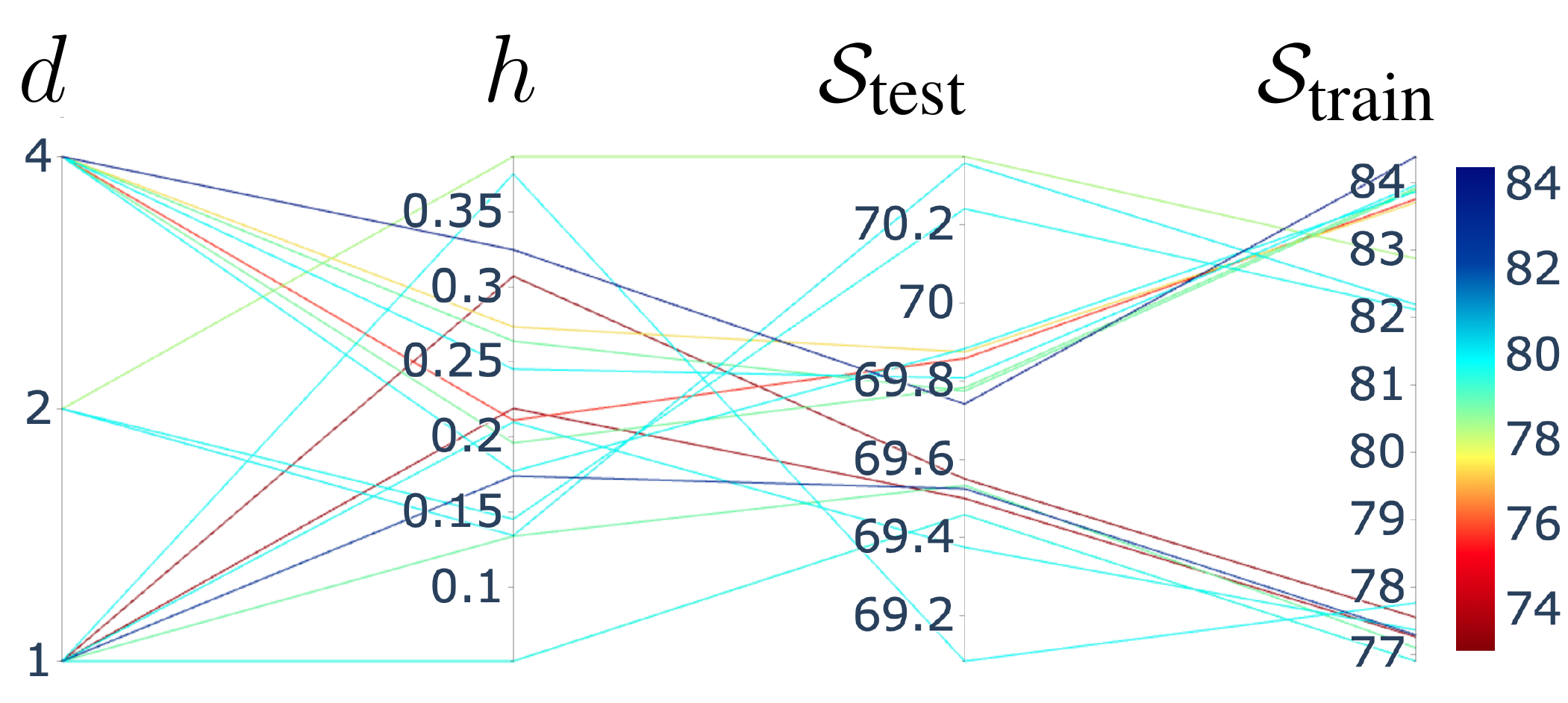}
    \caption{Results of hyperparameter tuning for the German credit dataset.}
        \label{fig:hope_german_dataset}
\end{figure}

Figure~\ref{fig:hope_german_dataset} shows the results of the hyperparameter tuning procedure over all sets of combinations of $d$ and $h$ for QMKL. The best resultant parameters are for the case of $d=2$ and $h=0.38$. The final results  presented in Table~\ref{tab:German_data} are based on these parameters used in the training of QMKL. For a fair comparison in terms of quantum resource usage, we use $d=2$ for SQKL and fixed-QMKL.

\end{document}